\documentclass[letterpaper, 10 pt, conference]{ieeeconf}
\IEEEoverridecommandlockouts   
\pdfoutput=1    
\usepackage{cite}
\usepackage{amsmath,amssymb,amsfonts}
\usepackage{algorithmic}
\usepackage{graphicx}
\usepackage{textcomp}
\usepackage{xcolor}
\usepackage{hyperref}
\usepackage{booktabs}
\usepackage{multirow}
\usepackage{tabularx}
\usepackage{courier}
\usepackage[font=footnotesize,skip=0pt]{caption}
\newcommand{\ra}[1]{\renewcommand{\arraystretch}{#1}}
\def\BibTeX{{\rm B\kern-.05em{\sc i\kern-.025em b}\kern-.08em
    T\kern-.1667em\lower.7ex\hbox{E}\kern-.125emX}}
\begin{document}

\title{Real-Time Predictive Control for Precision Machining
}


\author{Alexander Liniger, Luca Varano, Alisa Rupenyan and John Lygeros
\thanks{The authors are with the Automatic Control Laboratory, ETH Z\"urich, 8092 Z\"urich, Switzerland. A. Rupenyan is also with inspire AG, 8092 Z\"urich, Switzerland. (e-mails: 
\{liniger,ralisa,lygeros\}@control.ee.ethz.ch)} 
}

\maketitle

\begin{abstract}
Precise positioning and fast traversal times are crucial in achieving high productivity and scale in machining. This paper compares two optimization-based predictive control approaches that achieve high performance. In the first approach, the contour error is defined using the global position, the position on the path is inferred through a virtual path parameter, and the cost function combines the corresponding states and inputs to achieve a trade-off between high speed and positioning accuracy. The second approach is based on a local definition of both the error and the progress along the path, and results in a system with a reduced number of states and inputs that enables real-time optimization. Terminal and trust region constraints are required to achieve precise tracking of geometries where a fast or instantaneous change in direction is present. The performance of both approaches using different quadratic programming solvers is evaluated in simulations for geometries that are challenging in machine tools applications.

\end{abstract}


\section{Introduction}
Machine tool control of multi-axis systems is focused on accurate, high-speed tracking of a geometrical path \cite{MPCC,bx1,Huo2013}. The requirement for maximum productivity combined with high precision in contouring applications is reminiscent to the challenges encountered in autonomous race driving \cite{bx6,Liniger}. In high-precision cutting, the driver is the tool head, the lane width is the machining tolerance, and the optimized trajectory is the cutting contour. Similar methodology to generate a time-optimal trajectory can be applied, with the emphasis on tight tolerances in the order of tens of micrometers.
	
Model Predictive Contouring Control (MPCC) methods have been proposed  to increase the productivity of multi-axis computer numerical control (CNC) machine tools as they enable the coupled optimization of the velocity (feed) and the position of the tool. In \cite{MPCC} the contour, defined as the desired geometry to be traversed, is parametrized using the arc-length of the reference path. Based on the formulation presented in \cite{faulw}, where the cost function couples the contouring error with the progression on the geometrical path, a non-linear Model Predictive Control (MPC) formulation is proposed which trades off the contouring error and the traversal speed. The resulting non-linear MPC problem is solved by linearizing each segment, thus allowing to convert the problem into a quadratic program (QP). 


MPCC accounts for the real behavior of the machine and the axis drive dynamics can be excited to compensate for the contour error to a big extent, even without including friction effects in the model \cite{bx1, Stephens2013}. High-precision trajectories or set points can be generated prior to the actual machining process following various optimization methods, including MPC, feed-forward PID control strategies, or iterative-learning control \cite{Tang2013,bx5}, where friction or vibration-induced disturbances can be corrected. To achieve real-time performance with MPC, combined with accounting for persistent disturbances, the contouring error can be reduced by modifying the reference geometry offline based on the predicted contouring error \cite{bx2}. 
	
This work demonstrates two contouring control approaches, using MPC methods with a linear time-varying formulation. A modification of the MPCC method applied to biaxial machine tools  as implemented in \cite{MPCC} and \cite{MPCC_2} is compared with a local-variable method used in path following for autonomous driving and racing \cite{rajamani2011,novi2018,bx6,rucco2015efficient}. The two approaches differ in the definition of the contour-tracking error and how they tie it with the path. In the first approach the error is coupled with the progression along the path through the cost function. In the second approach the error is a component of the local coordinate transformation of the position along the path and the error progression is thus directly coupled to the system dynamics. The numerical implementation demonstrates on a simplified system excluding friction and oscillations that contour-tracking problems can be solved with a sampling rate in the order of 1 ms, making the methods suitable for real-time implementation.
\section{Problem Definition}
\label{prDef}
In this paper we look at a biaxial machine tool contouring control problem, where the goal is to traverse a given geometry as fast as possible while staying within a given tolerance band. We model the machine as a lumped mass model where the acceleration in $X$-$Y$ can be controlled individually. The resulting model is a classical double integrator model with the states given by $x = (X,Y,v_x,v_y)$ and the inputs given by $u = (a_x,a_y)$. The linear continuous time model can then be exactly discretized using periodic sampling and a Zero Order Hold (ZOH). The machine has independent acceleration limitations in the $X$-$Y$ direction of $\pm 20$\,m/s$^2$, resulting in the input constraints $u\in \mathcal{U} = \{u_x,u_y|\,|u_x| \leq 20\text{m/s}^2,\,|u_y| \leq 20\text{m/s}^2\}$. The velocity components $v_x, v_y$ are limited to $\pm 0.2$\,m/s, and  the velocity constraints are defined as $v \in \mathcal{V} = \{v_x,v_y| \, |v_x| \leq 0.2\text{m/s},\,|v_y| \leq 0.2\text{m/s}\}$. The geometry which should be traversed with the tool is parametrized by the arc-length of the curve $s \in [0,L]$ as path parameter: $\boldsymbol{r_d}(s)=(r_{d,x}(s), r_{d,y}(s))$. In our case the contour is given as piecewise linear in $s$, thus, the derivative of the geometry with respect to the $s$ can be computed and is given by $\boldsymbol{r'_d}(s)=(r'_{d,x}(s), r'_{d,y}(s))$ which can be interpreted as the tangent at the point $\boldsymbol{r_d}(s)$. Note that at the switching points one of the subgradients is used. The geometry also comes with a tolerance band the tool is not allowed to leave, in our case defined as $\pm 20$\,$\mu$m perpendicular to the contour. 
Based on these ingredients we can now formulate the MPC-based contouring control problem.

\section{Contouring control approaches}
\label{sec:approaches}
\subsection{Global variable MPCC}
\label{global}

The aim of the MPCC of \cite{MPCC} is to minimize the distance between the geometry and the optimized trajectory, while traversing the geometry as fast as possible \cite{Yuan2017}. As discussed in Section \ref{prDef}, the geometry is given by $\boldsymbol{r_d}(s)$ and is parametrized by the arc-length of the curve. The optimized trajectory at time $k$ is defined as $\boldsymbol{r_{d,k}} = (X_k,Y_k)$. A virtual path parameter $s_k$ is introduced as $s_{k+1}=s_{k}+T v_{s,k}$, where $v_{s,k}$ is the velocity along the path at time step $k$ and $T$ is the sampling time. The idea of the global variable MPCC formulation is to use the virtual path parameter $s_k$ to approximate the true path parameter $\hat s$ corresponding to a position $\boldsymbol{r_{d,k}}$. The difference in arc-length between $\boldsymbol{r_d}(\hat{s})$ and $\boldsymbol{r_d}(s_k)$ is defined as the lag error $e_{l,k}$, and the distance from $\boldsymbol{r_d}(\hat{s})$ to $\boldsymbol{r_{d,k}}$ as the contouring error $e_{c,k}$. Furthermore, the total error $\boldsymbol{r_d}(\hat{s})-\boldsymbol{r_{d,k}}$ is approximated with $\boldsymbol{r_d}(s_k)-\boldsymbol{r_{d,k}}$.  

\vspace{-1em}
\begin{figure}[htbp]
\label{fig:1}
\centerline{\includegraphics[width=6cm]{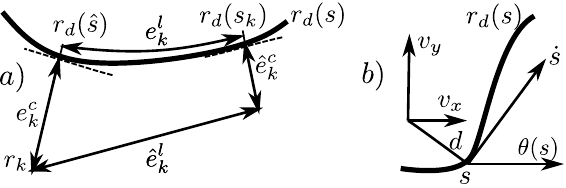}}
\caption{Definition of global variable (a) and local variable (b).}
\end{figure}
\vspace{-1em}

The resulting error vector is decomposed into a part $\hat{e}_{l,k}$  which is parallel to the tangent of the path at $s_k$ and another part $\hat{e}_{c,k}$ which is perpendicular to the tangent, the errors are defined as follows,

\vspace{-2em}
\begin{align}
\hat{e}_{l,k}(s_k) &=\frac{\boldsymbol{r'_d}(s_k)}{\|\boldsymbol{r'_d}(s_k)\|}\cdot( \boldsymbol{r_{d}}(s_k)-\boldsymbol{r_{d,k}})  \label{eq:1} \\
\hat{e}_{c,k}(s_k) &=\frac{\boldsymbol{r'_d}^\perp(s_k)}{\|\boldsymbol{r'_d}^\perp(s_k)\|}\cdot( \boldsymbol{r_d}(s_k)-\boldsymbol{r_{d,k}})  \, ,\label{eq:2}
\end{align}
\vspace{-1em}

where the parametric derivative $\boldsymbol{r'_d}$ is defined as $\boldsymbol{r'_d}(s)=(r'_{d,x}(s),r'_{d,y}(s))$ and the vector perpendicular to the tangent can be easily calculated as  $\boldsymbol{r'_d}^\perp(s)=(-r'_{d,y}(s),r'_{d,x}(s))$.

If $\hat{e}_{l,k}$ is small, $\hat{e}_{c,k}$ is a good approximation of the contour error and the virtual path parameter is a good approximation of $\hat{s}$. The virtual parameter represents the progression on the path which can be controlled with the input $v_{s,k}$. The connection between the longitudinal error and the path parameter is a key feature of the MPCC which ties the progression on the path with the contour error and is later included into the cost function. 

The errors as defined in \eqref{eq:1} and \eqref{eq:2} solely depend on states at time $k$. However, fast QP solvers such as the ones used in our simulation study, do only allow for equality constraints linking two consecutive time steps. Thus, we reformulate the errors at time step $k+1$ to depend only on information of time step $k$, thus introducing error dynamics. More precisely the error dynamics depends on the state $x_k$ and inputs $u_k$ of the lumped mass model, as well as on the errors $e_{l,k},e_{c,k}$, the virtual path parameter $s_k$, and the velocity along the path $v_{s,k}$. As a first step the error is linearly approximated along the geometry around $s_k$:

\vspace{-1em}
\begin{align} 
\hat{e}_{l,k+1}&=\frac{{r'_{d,x}}(s_k)}{\| \boldsymbol{r'}(s_k) \|}(r_{d,x}(s_k)-X_{k+1}) \nonumber \\&+\frac{{r'_{d,y}}(s_k)}{\| \boldsymbol{r'}(s_k) \|}(r_{d,y}(s_k)-Y_{k+1})+T v_{s,k} \label{eq:3} \\
\hat{e}_{c,k+1}&=-\frac{{r'_{d,y}}(s_k)}{\| \boldsymbol{r'}(s_k) \|}(r_{d,x}(s_k)-X_{k+1}) \nonumber \\&+\frac{{r'_{d,x}}(s_k)}{\| \boldsymbol{r'}(s_k) \|}(r_{d,y}(s_k)-Y_{k+1})  \, . \label{eq:4}
\end{align}	
\vspace{-1em}

The $k+1$ terms on the right hand side can be replaced with terms known from the lumped mass dynamics: $X_{k+1} = X_k + T v_{x,k} + T^2/2\, a_{x,k}$, $Y_{k+1} = Y_k + T v_{y,k} + T^2/2 \,a_{y,k}$. The resulting larger dynamical system has a state given by $\hat{x} = (X,Y,v_x,v_y,s,e_l,e_c)$ and an input $\hat{u} = (a_x,a_y,v_s)$, but, due to \eqref{eq:3} and \eqref{eq:4} the system is no longer linear. The non-linearity comes from the geometry terms $\boldsymbol{r}(s)$ and $\boldsymbol{r'}(s)$, which are non-linear in $s$. Since we solve the problem in a receding horizon fashion, we can use the shifted previous solution of the $s$ state as a guess for the solution and linearize the geometry terms around this estimated $s$ trajectory. As long as the solutions between consecutive MPC solutions do not differ too much, this approach should result in good approximations of the error dynamics. This linearization results in a linear time varying system of the form $\hat{x}_{k+1} = \hat{A}_k \hat{x}_k + \hat{B}_k \hat{u}_k + \hat{g}_k$, where only the error dynamics are time-dependant.

Finally, we design a cost function that matches the goals of the contouring controller. We include a term penalizing the squared longitudinal error $\hat{e}_l^2$ since this error has to be small for the formulation to be accurate, and a term to minimize the squared contouring cost $\hat{e}_{c}^2$ as this represents our goal of following the geometry closely. We reward progress at the end of the horizon $s_N$, which corresponds to traversing the geometry as fast as possible, and penalize the squared velocities and applied inputs, to have smooth velocity and input trajectories. The resulting MPCC problem has the following form,

\vspace{-1em}
\begin{align}
\min_{\mathbf{\hat{x}},\mathbf{\hat{u}}}\quad &\sum_{k=1}^{N-1}  \gamma_{l} \hat{e}_{l,k}^{\:2} +   \gamma_{c} \hat{e}_{c,k}^{\:2} + v_k^T Q_v v_k + u_k^T R u_k  \nonumber\\&+\gamma_{l,T} \hat{e}_{l,N}^{\:2} +\gamma_{c,T} 
\hat{e}_{c,N}^{\:2} + v_N^T P_{v} v_N   -  \gamma_{s} s_N \nonumber\\
\text{s.t} \quad\; & \hat{x} = \hat{x}(0)\nonumber\\
&\hat{x}_{k+1} = \hat{A}_k \hat{x}_k + \hat{B}_k \hat{u}_k + \hat{g}_k\nonumber\\
&\hat{e}_{c,k} \in \mathcal{T}^c, \quad v_{k} \in \mathcal{V}, \quad u_k \in \mathcal{U} \nonumber\\
&v_{N} \in \mathcal{V}_T,\:\hat{e}_{c,N} \in \mathcal{T}^c_{T} \nonumber\\
&k=0,..,N-1 \label{eq:globalMPC}
\end{align}
\vspace{-1em}

where $\mathbf{\hat{x}} = (\hat{x}_0,...,\hat{x}_N)$ and $\mathbf{\hat{u}} = (\hat{u}_0,...,\hat{u}_{N-1})$ are the state and input trajectories. $\gamma_{l}$ and $\gamma_{c}$ are the error weights, $Q_v$ and $R$ are positive definite velocity and input weight matrices. The terminal cost consists of the lag, contouring, and velocity weights $\gamma_{l,T}$, $\gamma_{c,T}$ and $P_v$, as well as the progress maximization weight $\gamma_s$. The MPCC problem constrains the contouring error to stay within the tolerance band, which we denote by $\mathcal{T}^c$, in addition to the velocity and input constraints mentioned in Section \ref{prDef}. Finally, we impose terminal constraints which constrain the velocity to $\pm 0.002$ m/s and the contouring error to $\pm 20$ $\mu$m. The terminal cost and constraints are imposed to deal with potentially fast changing geometries not yet ``seen" by the MPC, which would otherwise result in recursive feasibility issues. We discuss the implications of the terminal constraints further in Section \ref{sec:tightCorners}.

\subsection{Local Variable Approach}
\label{subsec:local}
	The global variable approach uses the global position to define the errors and a virtual path parameter to define the position on the path. The error states are recomputed at every time step and are only introduced into the dynamics such that a cost and constraints can be assigned to them, resulting in a system with some redundant states. To simplify the problem, a second approach is implemented, where a local definition for the error and the progression on the path is used, resulting in a state space system with a reduced number of states and inputs, all having real dynamics. This implementation is inspired by path following controllers in autonomous driving such as the methods proposed in \cite{rajamani2011}.  
The idea is to describe the system in a local curvilinear coordinate system, where the local state is formed by the velocities, the path parameter $s$, and the perpendicular distance from the path to the machine position, which we call $d$ (see Fig 1b). Note that given these coordinates and the path, the global coordinates can be reconstructed. The dynamics in this local coordinate system can be formulated given the local angle of the path, which is commutable by the parametric derivative $\boldsymbol{r'_d(s)}$ as $\theta(s) = \text{atan2}(r'_{d,y}(s),r'_{d,x}(s))$. The movement along the path can then be described through the projection of the horizontal and the vertical velocities on the path,
	\begin{equation}
     \dot{s}=\frac{v_x\cos(\theta(s))+v_y\sin(\theta(s))}{1-\kappa(s) d} ,
     \label{s_dot}
	\end{equation}
where $\kappa(s)$ is the local curvature. For the geometries and tracking errors that arise in this application the denominator in \eqref{s_dot} is roughly equal to 1. We will therefore ignore the dependence on the curvature in the sequel. To simplify the notation, we will also drop the dependence of the angle on $s$ and write simply $\theta$ in place of $\theta(s)$. Similar to the dynamics along the path, the movement perpendicular to the path is the projection on the vector perpendicular to the tangent,
	\begin{equation}
	\dot{d}=-v_x\sin(\theta)+v_y\cos(\theta)  \, .
	\end{equation}
	The resulting state of the system is given by $\tilde{x} = (v_x,v_y,s,d)$ and the input is again $\tilde{u} = (a_x,a_y)$. Following ZOH discretization the system dynamics is given by:
	\vspace{-1em}
	\begin{equation}
	\begin{aligned}
	\begin{bmatrix}
	v_{x,k+1} \\
	v_{y,k+1} \\
	s_{k+1} \\
	d_{k+1} 
	\end{bmatrix}
	=&
	\begin{bmatrix}
	1 &0 &0 &0& \\
	0 &1 &0 &0& \\
	\cos(\theta)T &\sin(\theta)T &1 &0& \\
	-\sin(\theta)T &\cos(\theta)T &0 &1& 
	\end{bmatrix}
	\begin{bmatrix}
	v_{x,k} \\
	v_{y,k} \\
	s_k \\
	d_k 
	\end{bmatrix}
	+ \\
	&
	\begin{bmatrix}
	T &0\\
	0 &T\\ 
	\cos(\theta)T^2/2 &\sin(\theta)T^2/2\\
	-\sin(\theta)T^2/2 &\cos(\theta)T^2/2
	\end{bmatrix}
	\begin{bmatrix}
	a_{x,k} \\
	a_{y,k} 
	\end{bmatrix}  \, .
	\end{aligned} \label{eq:localDyn}
	\end{equation}
Since the angle $\theta$ is a non-linear function of the path parameter $s$, we again use the solution of the previous MPC problem to linearize these non-linear terms, as in the global variable approach \ref{global}.
	
The resulting cost function only needs a weight to minimize the deviation from the path $d$, penalization of the velocities and inputs for smooth trajectories, and a reward on the path progression $s_N$. Altogether the following local MPC problem can be formulated,	
%
%
\vspace{-1em}
\begin{align}
\min_{\mathbf{\tilde{x}},\mathbf{\tilde{u}}} \; &\sum_{k=1}^{N-1}  \gamma_{d} d_k^2 + v_k^T Q_v v_k + u_k^T R u_k  \nonumber\\& +\gamma_{d,T} d_N^2 + v_N^T P_{v} v_N   -  \gamma_{s} s_N\nonumber\\
\text{s.t} \quad\; & \tilde{x} = \tilde{x}(0)\nonumber\\
&\tilde{x}_{k+1} = \tilde{A}_k \tilde{x}_k + \tilde{B}_k \tilde{u}_k \nonumber\\
&d_{k} \in \mathcal{T}^c, \quad v_{k} \in \mathcal{V}, \quad u_k \in \mathcal{U} \nonumber\\
&v_{N} \in \mathcal{V}_T,\:d_{N} \in \mathcal{T}^c_{T} \nonumber\\
&k=0,..,N-1 \label{eq:localMPC}
\end{align}
where, $\mathbf{\tilde{x}} = (\tilde{x}_0,...,\tilde{x}_N)$ and $\mathbf{\tilde{u}} = (\tilde{u}_0,...,\tilde{u}_{N-1})$ are the state and input trajectories. $\gamma_{d}$ is the weight on the deviation from the path, $Q_v$ and $R$ are positive definite velocity and input weight matrices. Similar to the global MPC problem \eqref{eq:globalMPC}, the terminal cost consists of higher contouring and velocity weights $\gamma_{d,T}$ and $P_v$, as well as the progress maximization weight $\gamma_s$. The constraints as well as the terminal constraints are identical to those of the global MPC problem \eqref{eq:globalMPC}, with the only difference that the deviation from the path is denoted by $d$ instead of $e_c$.

\subsection{Dealing with sharp corners}\label{sec:tightCorners}

\subsubsection{Terminal Ingredients}
The terminal cost and constraints are of fundamental importance for this application, since the tool should be able to traverse sharp corners that  may result in the MPC optimization problem becoming infeasible. This can be avoided by requiring that the tool has to be able to stop on the reference path at the end of the horizon, as from such a state any subsequent geometry can be traversed. A zero velocity terminal constraint ($v_{x,N} = v_{y,N} = 0$) does even guarantee recursive feasibility of the problem as for both systems this is an equilibrium. In our implementation we use a relaxed version of the constraint as some of the used QP solvers do not allow for zero velocity terminal constraints, where $|v_{x,N}| <= 0.002$ m/s and $|v_{y,N}| <= 0.002$ m/s, combined with a high terminal quadratic cost on the velocities. 
The addition of these terminal constraints may result in conservative performance for short horizons, as they implicitly limit the maximum velocity to a velocity where the tool can decelerate to a standstill. For longer horizon the influence on the closed loop results becomes negligible.

\subsubsection{Trust Region Constraints}
The local MPC formulation model heavily depends on the angles $\theta (s)$, however, as the trajectory changes slightly form iteration to iteration, the linearization points for these angles are not completely correct. This can result in large model prediction errors, especially if the tool needs to traverse corners with small radius (\textless1mm) or sharp corners. Therefore, we impose trust region constraints for $s_k$ to force the current solution to remain close to the previous solution. This results in low prediction errors and is essential for successful traversal of tight corners as presented in the numerical results. The corresponding additional constraints to \eqref{eq:localMPC} are of the form $\underline{s}_k \leq s_k \leq \overline{s}_k$, where to $\underline{s}_k$ and $\overline{s}_k$ depend on the previous MPC solution. Note that every time step has its own bounds. 

\subsubsection{Local MPCC State Feedback}
Due to the model mismatch introduced by the changing angle of the geometry changing over one time step, simply simulating the dynamics \eqref{eq:localDyn} is not a valid option. This  is especially true for sharp corners where the angle changes rapidly (in the limit discontinuously). Therefore, we instead simulate the lumped mass model in global coordinates and then project the position onto the path. Since we have a good initial guess for the location of the projection, we can find locally the closest segment of our piecewise linear geometry and then project onto this segment using an inner product. Note that this is not necessary for the global MPCC approach as the system is formulated in global coordinates.

\section{Results} \label{sec:results}
The simulations are executed in Arch Linux on a desktop PC with an Intel Core i9 9900K CPU. 	
All files are compiled using gcc with the -O3 option enabled. We have implemented the simulations on \texttt{acados}, an interface to fast and embedded solvers for nonlinear optimal control and dynamic optimization \cite{acados}, providing a convenient framework to evaluate various MPC implementations and solver performance. Currently available MPC solvers are qpOASES \cite{qpoases}, HPIPM using BLASFEO \cite{blasfeo}, and HPMPC \cite{hpmpc}. In all simulations the system is modelled using the lumped mass double integrator model.

 The performance of the two MPCC approaches is assessed by the RMS error in tracking of the target geometry, the infinity-norm tracking error, and by the maneuver time (the time needed for the tool to complete the geometry). We have investigated the effect of modifying the penalty on the contouring error in the cost function which controls the trade-off between tracking accuracy and machining time. The sampling time of the MPC is 1 ms and the horizon length is chosen such that the terminal constraints do not affect the performance. The performance of the two approaches is compared for horizons of 35 and 70 time steps using the HPIPM solver. We first investigate the tracking performance of the two approaches on a smooth geometry, then on a geometry with sharp edges, and finally compare the computational performance and the available QP solvers.
\subsection{Smooth geometry}\label{sectionRounded}
\begin{figure}[htbp]
\centerline{\includegraphics[width=8cm]{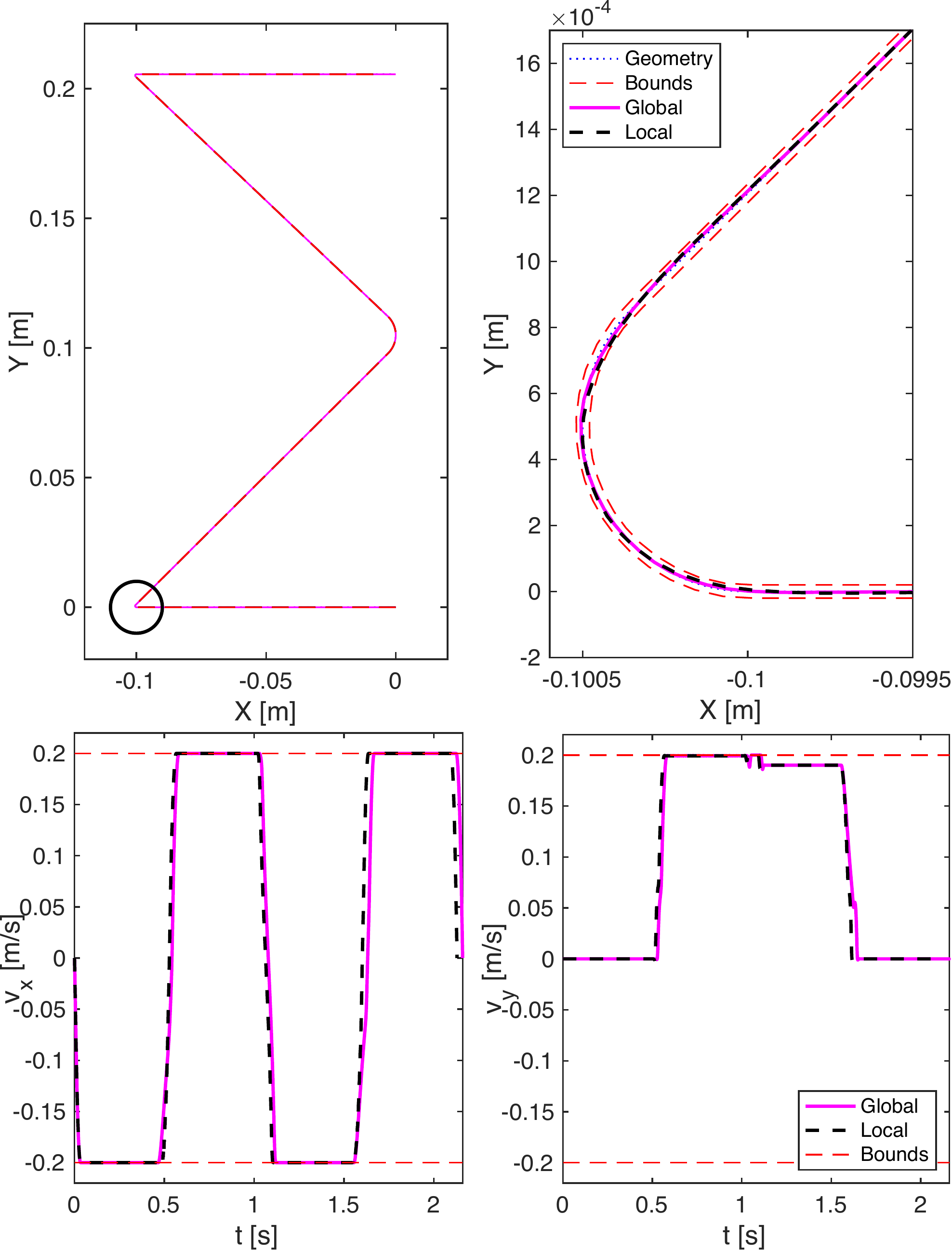}}
\caption{Experimental results for the smooth geometry using long horizon and high contouring cost.}
\label{fig:3}
\vspace{-1em}
\end{figure}

To assess the effect of large geometric variations on the performance of the two approaches, we have tested the two approaches on two similar paths based on the Greek letter $\Sigma$, one with sharp and one with rounded corners. The sigma geometry is 10 cm wide and 20 cm high, for both geometries the middle edge is rounded and has a radius of 1 cm (see Figure \ref{fig:3} and \ref{fig:4}). The other two corners are different for the two geometries, where the smooth geometry has rounded corners with a radius of 0.5 mm (see Figure \ref{fig:3}), and the geometry with the sharp corners has an instantaneous change in direction (see Figure \ref{fig:4}). 

The weights of both the global and local MPCC were set to be equal to compare the results, and only the contouring error weight and the horizon length were changed.
Figure \ref{fig:3} shows the experimental results for the geometry with rounded corners with long horizon length and high contouring cost, and Table \ref{tab:performanceRound} summarizes the tracking error (root mean square (RMS)-tracking and infinity-norm tracking) and the maneuver time for each combination of contouring cost and horizon length.
\begin{table}
\caption{Performance smooth geometry}
\label{tab:performanceRound}
\centering 
\ra{1.3}
\begin{tabular}[h]{@{}l c c c c c @{}}\toprule
& \multicolumn{2}{c}{$N = 35$} & &  \multicolumn{2}{c}{$N = 70$}\\
\cmidrule{2-3} \cmidrule{5-6}
 & global &  local & & global & local \\
 \midrule
\textit{high contouring cost} \\ 
RMS-tracking [$\mu$m]   & 0.472 & 0.823 & & 0.821 & 1.054 \\
inf-norm tracking [$\mu$m] & 5.701 & 14.336 & & 13.675 & 13.775\\
Maneuver time [s]   & 2.438 & 2.400 & & 2.161 & 2.130 \\ 
\midrule
\textit{low contouring cost} \\
RMS-tracking [$\mu$m]   & 7.176 & 5.377 & & 14.089 & 13.868 \\
inf-norm tracking [$\mu$m] & 20.000 & 20.000 & & 20.000 & 20.000\\
Maneuver time [s]       & 2.430 & 2.397 & & 2.150 & 2.129 \\
\bottomrule
\end{tabular}
\end{table} 
For the high contouring cost case ($\gamma_c = \gamma_d = 10^8$), the tracking error increases for long horizon lengths ($N = 70$) compared to the short horizon lengths ($N = 35$), and the maneuver time decreases with about 10\%. The decrease in accuracy is two-fold for the global variable approach, whereas for the local variable it is only 10-15$\%$, with comparable decrease of the manoeuvre time for both.

The main difference between the long and the short horizon is the speed on the diagonal straight pieces. There the velocity is lower for the short horizon, primarily due to the terminal ingredients. Removing the terminal ingredients allows for higher velocities in these segments, at the expense of the theoretical guarantees provided by the terminal constraints. The velocity can be increased further by using more aggressive weights (increased $\gamma_s$), since the horizon is in theory long enough to come to standstill from initial velocity. However, more aggressive weights result in controllers not being able to complete the whole geometry. This effect is present in both geometries and both controllers, but the global variable MPCC is less effected. 
For the long horizon the terminal cost and constraints have no influence on the performance and the controller, leading to faster maneuver times. However, the longer horizon also exploits possible cutting of corners leading to an increased tracking error.

For the low contouring cost case ($\gamma_c = \gamma_d = 1$), as expected the tracking error is significantly higher, and for all four cases the controller reaches the limit of the tolerance band. Similar to the high cost case the local variable MPCC is slightly faster for the same parameters. However, the maneuver time is only marginally shorter for the low contouring case than for the high contouring case, which leads to the conclusion that a low contouring cost is not preferable for our application.

\subsection{Sharp corner geometry}
\label{sec:sharp}
\begin{figure}[htbp]
\centerline{\includegraphics[width=8cm]{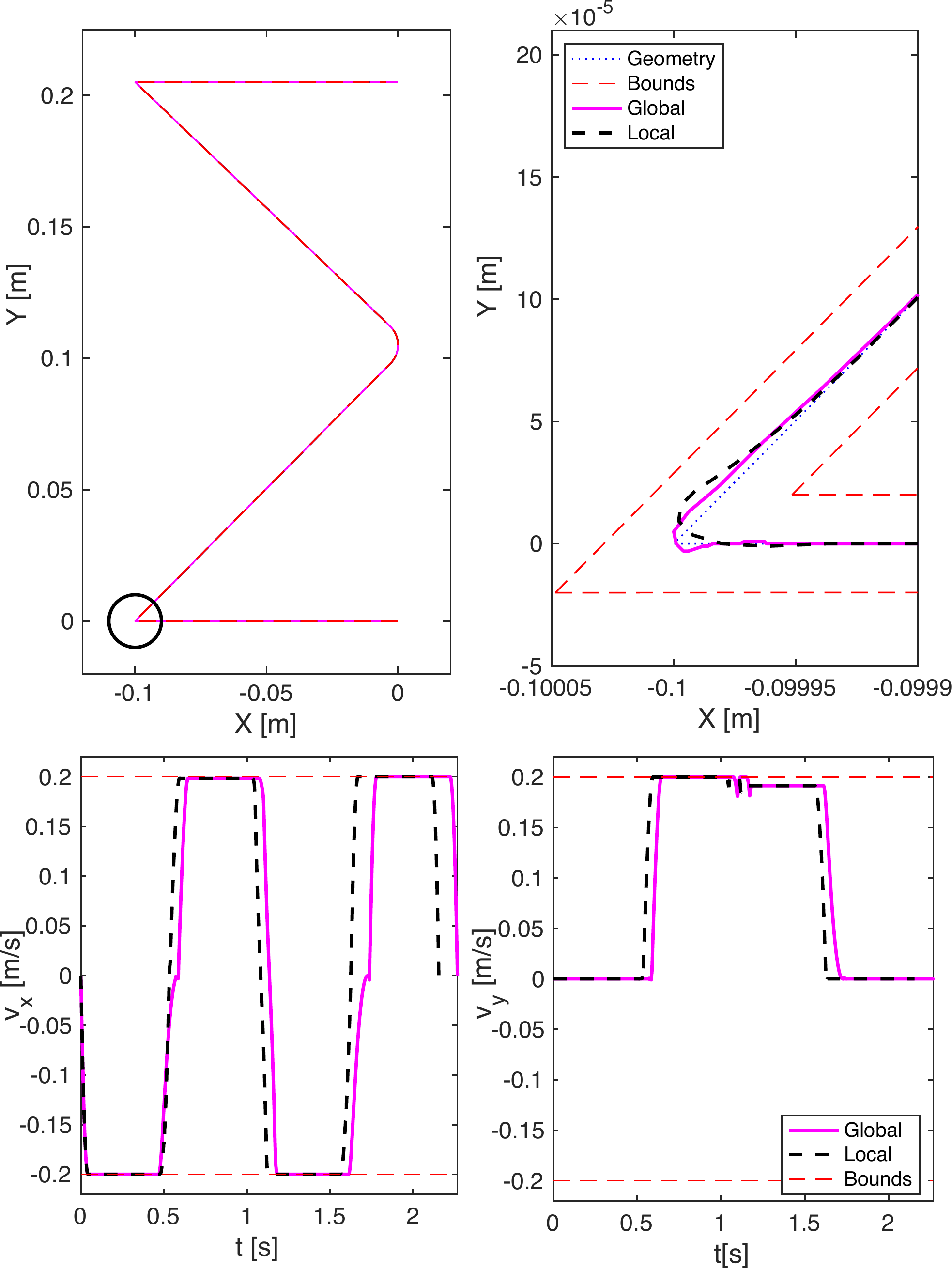}}
\caption{Experimental results for the sharp corner geometry using long horizon and high contouring cost.}
\label{fig:4}
\end{figure}

Figure \ref{fig:4} shows the experimental results for the geometry with sharp corners, with a long horizon and high contouring cost. The contouring error (RMS-tracking and infinity norm tracking) and the maneuver time for high contouring cost and horizon lengths of 35 and 70 are summarized in Table \ref{tab:N_S2}. The same trends as in the smooth geometry simulations are present. The tracking accuracy is higher for short horizons, leading to increased maneuver times. The local variable approach in all cases results in a slightly faster maneuver times, and in increased tracking error. The increased tracking error of the local variable method again is caused by the model mismatch introduced when traversing the corner. Compared to the smooth geometry the maneuver time is slower, since only a less aggressive controller was able to traverse the edge. Also note that for the sharp corner geometry only the high contouring cost successfully finished the geometry, without getting stuck at infeasible points. However, the tracking is improved, which is due to the corner being just one point, resulting in less room where the tool should deviate from the path.
\begin{table}
\caption{Performance sharp corner geometry}
\label{tab:N_S2}
\centering 
\ra{1.3}
\begin{tabular}[h]{@{}l c c c c c @{}}\toprule
& \multicolumn{2}{c}{$N = 35$} & &  \multicolumn{2}{c}{$N = 70$}\\
\cmidrule{2-3} \cmidrule{5-6}
 & global &  local & & global & local \\
 \midrule
\textit{high contouring cost} \\ 
RMS-tracking [$\mu$m]   & 0.190 & 0.717 & & 0.373 & 0.631 \\
inf-norm tracking [$\mu$m] & 3.142 & 15.914 & & 5.036 & 10.208\\
Maneuver time [s]   & 3.632 & 3.600 & & 2.267 & 2.155 \\ 
\bottomrule
\vspace{-1em}
\end{tabular}
\end{table}
\vspace{-1em}	
\vspace{-1em}

	
The velocity profiles for the rounded corners (Figure \ref{fig:3}) and the sharp corners (Figure \ref{fig:4}) are similar. However, in the case of the sharp corner the controller slows down more to traverse the edge, especially in the case of the global variable MPCC, where the tool nearly comes to a halt. 
\subsection{Solver performance}
To assess the performance of the solvers, we focus on the smooth geometry with high contouring cost case. Note that the results for the other cases are very similar. We compare the performance of the solvers HPIPM, HPMPC and qpOASES in terms of the average and maximum computation time. As usual the maximum computation time can be influenced by other factors, depending on the processing power and configuration. 

Table \ref{tab:solver} shows that the global variable MPCC approach is about 2-2.5 times slower than the local variable MPCC. This is expected as the combined number of states and inputs is reduced from 10 to 6 with the local variable approach, which significantly reduces the number of optimization variables. Note that qpOASES was not able to solve the global variable MPCC approach, whereas the local approach could be solved successfully. When comparing the solvers for the local variable MPCC, Table \ref{tab:solver} shows that qpOASES is the fastest solver for short horizon length, while HPMPC is fastest for the long horizon length. For long horizons it can be clearly seen that the computation time of qpOASES increases to 10 times the computation time with short horizon, whereas the time for HPIPM and HPMPC doubles. This is expected, as HPIPM and HPMPC are tailored sparsity exploiting MPC solvers, where the complexity grows linear in the horizon length. On the other hand, qpOASES requires a dense condensed MPC problem resulting in a cubic complexity in the horizon length. For the global variable approach where qpOASES did not solve the problem, HPMPC was the fastest solver for both horizon lengths. Note that all computation times include setting up the QP. Even though HPIMP is the slowest solver of the three, the performance is still impressive and the solver includes features not available in HPMPC.
\begin{table}
\caption{Computation times smooth geometry}
\label{tab:solver}
\centering 
\ra{1.3}
\begin{tabular}[h]{@{}l c c c c c @{}}\toprule
& \multicolumn{2}{c}{$N = 35$} & &  \multicolumn{2}{c}{$N = 70$}\\
\cmidrule{2-3} \cmidrule{5-6}
 & global &  local & & global & local \\
 \midrule
\textit{HPIPM} \\ 
mean [ms]   & 1.843 & 0.709 & & 3.874 & 1.533 \\
max [ms]   & 2.654 & 1.087 & & 5.734 & 2.200\\ 
\midrule
\textit{HPMPC} \\ 
mean [ms]   & 0.849 & 0.431 & & 1.949 & 0.901 \\
max [ms]   &  1.145 & 0.620 & & 2.978 & 1.187 \\ 
\midrule
\textit{qpOASES} \\ 
mean [ms]   & - & 0.295 & & - & 2.938 \\
max [ms]   & - & 0.849 & & - & 12.479 \\ 
\bottomrule
\vspace{-2em}
\end{tabular}
\end{table}
For short horizons and the local variable MPCC all solvers reach a computation time lower than 1 ms, suitable for real-time implementation. The maximum computation time with HPMPC is 0.62 ms, making it suitable for further on-machine implementation and testing on the experimental set up.
	\section{Conclusion and Outlook} \label{sec:conclusion}
In this paper, we presented two formulations for contour tracking problems, using model predictive control with a QP solver implementation on a simplified system, excluding non-linear effects. The performance of the two formulations was investigated for different horizon lengths, for a smooth geometry and a sharp corner geometry, where tracking is constrained within a tolerance band of $\pm 20$ $\mu$m.
	Both global and local variable MPCC approaches achieve accurate tracking of the two target paths, even in the more challenging case of the sharp corners geometry. The MPCC for a biaxial stage could be successfully implemented with sub-ms computation times using the local variable approach, both on smooth geometries and geometries with sharp corners. Its good performance in computation time and geometry tracking make it a good candidate for industrial applications.  The presented simulations exclude friction and oscillatory behavior. Once the response of the tool following a given geometry is known, it can be included with a tracking MPC formulation. The resulting increase in computation time due to the increased number of states could still be accommodated using the local coordinates approach, which is a focus of future research. 
\bibliography{IEEEabrv,MPCC_bib.bib} 
\bibliographystyle{IEEEtran}

\end{document}